\documentclass[conference]{IEEEtran}
\PassOptionsToPackage{table}{xcolor}

\usepackage[style=numeric-comp, maxnames=2, minnames=1, abbreviate=true, dateabbrev=true, alldates=year, doi=true, isbn=false, url=false, eprint=false, language=american, giveninits=true, sortcites=true, sorting=none, backend=biber]{biblatex} 

\AtEveryBibitem{\clearlist{location}} %
\AtEveryBibitem{\clearlist{language}} %
\AtEveryBibitem{\clearfield{eprintclass}} %

\newcommand{\changed}[1]{\textcolor{black}{#1}}
\usepackage{caption}
\usepackage{flushend}
\usepackage{framed}
\usepackage{svg}
\usepackage{listings}  
\usepackage[caption=false]{subfig}
\usepackage{mathtools}
\PassOptionsToPackage{hyphens}{url}
\usepackage{hyperref}
\hypersetup{ %
    colorlinks,
    linkcolor={red!40!black},
    citecolor={blue!40!black},
    urlcolor={blue!40!black}
}
\usepackage{adjustbox}
\usepackage{enumitem}

\usepackage{makecell}
\usepackage{tcolorbox}
\usepackage{todonotes}
\usepackage{booktabs}
\usepackage{multirow}
\usepackage{xcolor}
\usepackage{xspace}
\usepackage{tikz}
\def\checkmark{\tikz\fill[scale=0.4](0,.35) -- (.25,0) -- (1,.7) -- (.25,.15) -- cycle;} 

\usepackage{listings}
\definecolor{codegreen}{rgb}{0,0.6,0}
\definecolor{codegray}{rgb}{0.5,0.5,0.5}
\definecolor{codepurple}{rgb}{0.58,0,0.82}
\definecolor{backcolour}{rgb}{0.95,0.95,0.92}

\lstdefinestyle{mycode}{
    backgroundcolor=\color{backcolour},   
    commentstyle=\color{codegreen},
    keywordstyle=\color{magenta},
    numberstyle=\tiny\color{codegray},
    stringstyle=\color{codepurple},
    basicstyle=\ttfamily\footnotesize,
    breakatwhitespace=false,         
    breaklines=true,                 
    captionpos=b,                    
    keepspaces=true,                 
    numbers=left,                    
    numbersep=5pt,                  
    showspaces=false,                
    showstringspaces=false,
    showtabs=false,                  
    tabsize=2
}

\definecolor{Gray}{gray}{0.95}
\definecolor{Gray2}{gray}{0.92}

\usepackage{fancybox}
\usepackage[framemethod=tikz]{mdframed}
\mdfdefinestyle{mystyle}{%
  rightline=false,
  innerleftmargin=5,
  innerrightmargin=5,
  topline=false,
  leftline=false,
  bottomline=false,
  skipabove=\topsep,
  skipbelow=\topsep,
  backgroundcolor = lightgray,
}

\newcommand{\appliedTransforms}{16\xspace} 
\newcommand{\pubSemantic}{28\xspace} 
\newcommand{\pubRelevant}{19\xspace} 

\newcommand{\numTransforms}{94\xspace}
\newcommand{\head}[1]{\par\noindent\textbf{#1:}\space}

\begin{document}

\title{Semantic-Preserving Transformations as Mutation Operators: A Study on Their Effectiveness in Defect Detection}

\author{\IEEEauthorblockN{Max Hort}
\IEEEauthorblockA{Simula Research Laboratory\\
Oslo, Norway\\
Email: maxh@simula.no}
\and
\IEEEauthorblockN{Linas Vidziunas}
\IEEEauthorblockA{Simula Research Laboratory\\
Oslo, Norway\\
Email: linasvidz@simula.no}
\and
\IEEEauthorblockN{Leon Moonen}
\IEEEauthorblockA{Simula Research Laboratory\\
Oslo, Norway\\
Email: leon.moonen@computer.org}
}

\thispagestyle{plain}
\pagestyle{plain}

\makeatletter
\def\ps@IEEEtitlepagestyle{%
  \def\@oddfoot{\mycopyrightnotice}%
  \def\@evenfoot{}%
}
\def\mycopyrightnotice{%
  \hspace*{3mm}\includegraphics[width=2cm]{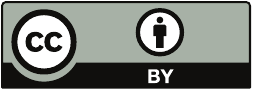}%
  \hspace*{2mm}\raisebox{2.5mm}{%
          \parbox{\columnwidth}{\footnotesize This work is licensed under a Creative Commons \\ Attribution 4.0 International (CC BY 4.0) license.}%
          \hspace*{-73pt}\mbox{\thepage}\hspace{10pt}\fbox{\parbox{.94\columnwidth}{\footnotesize\textsl{Accepted for publication in Mutation 2025 at the 18th IEEE International Conference on Software Testing, Verification and Validation (ICST 2025).}}}%
  }%
  \gdef\mycopyrightnotice{}%
}
\makeatother

\clearpage{}%

\maketitle

\begin{abstract}
Recent advances in defect detection use language models. \changed{Existing works enhanced the training data to improve the models' robustness when applied to semantically identical code (i.e., predictions should be the same). However, the use of semantically identical code has not been considered for improving the tools during their application - a concept closely related to metamorphic testing.}

The goal of our study is to determine whether we can use semantic-preserving transformations, analogue to mutation operators, to improve the performance of defect detection tools in the testing stage.
We first collect existing publications which implemented semantic-preserving transformations and share their implementation, such that we can reuse them.
We empirically study the effectiveness of three different ensemble strategies for enhancing defect detection tools.
We apply the collected transformations on the Devign dataset, considering vulnerabilities as a type of defect, and two fine-tuned large language models for defect detection (VulBERTa, PLBART). 

We found \pubSemantic publications with \numTransforms different transformation. We choose to implement 39 transformations from four of the publications, but a manual check revealed that 23 out 39 transformations \textbf{change code semantics}.
\changed{Using the \appliedTransforms remaining, correct transformations} and three ensemble strategies, we were not able to increase the accuracy of the defect detection models.
Our results show that reusing shared semantic-preserving transformation is difficult, sometimes even causing wrongful changes to the semantics. 
\end{abstract}

\begin{IEEEkeywords}
defect detection, language model, semantic-preserving transformation, ensemble
\end{IEEEkeywords}

\section{Introduction}
\noindent
Among others, defect detection models should be accurate (\textit{good at detecting defective code snippets}) and robust (\textit{making the same prediction for functions that are doing the same thing}).
Existing research measured how robust defect detection models are~\cite{pour2021:searchbased} and improved robustness via the training process (e.g., adversarial training) by incorporating code snippets that look different but have the same functionality~\cite{pour2021:searchbased,jain2021:contrastive}. 
One approach to obtain such snippets is the use semantic-preserving transformations (e.g., changing a for loop to a while loop), which do not change the code snippets' functionality.
An unexplored aspect is whether ``unrobust'' models can be improved by using the diverse predictions made on code transformed with semantic-preserving transformations.

This intuition is inspired by metamorphic testing, which instead of ground truth datasets, has access to inputs which should generate the same output~\cite{applis2021:assessing,applis2023:searching}. 
Models are then evaluated by their ability to make the same predictions for such input pairs.
Rather than comparing the predictions made on multiple inputs, as done in metamorphic testing, we use them to improve the performance of LLMs for defect detection.

As an example, imagine you have written code which a defect detection model predicts as ``correct''. You now go ahead and change a variable name. All of a sudden the prediction changes to defective. Surprised, you repeat the same procedure another 5 times but the prediction remains as defective. 
\changed{This raises the question whether to trust the first prediction, as it was performed on the original code snippet or the majority of tests, which were labeled as defective.}
Following this intuition, we study the following:
\textbf{What is the impact of semantic-preserving transformations on defect detection tool performance?}

\begin{figure*}[!htb]
\vspace{2mm}
\centering
\includegraphics[width=.95\textwidth]{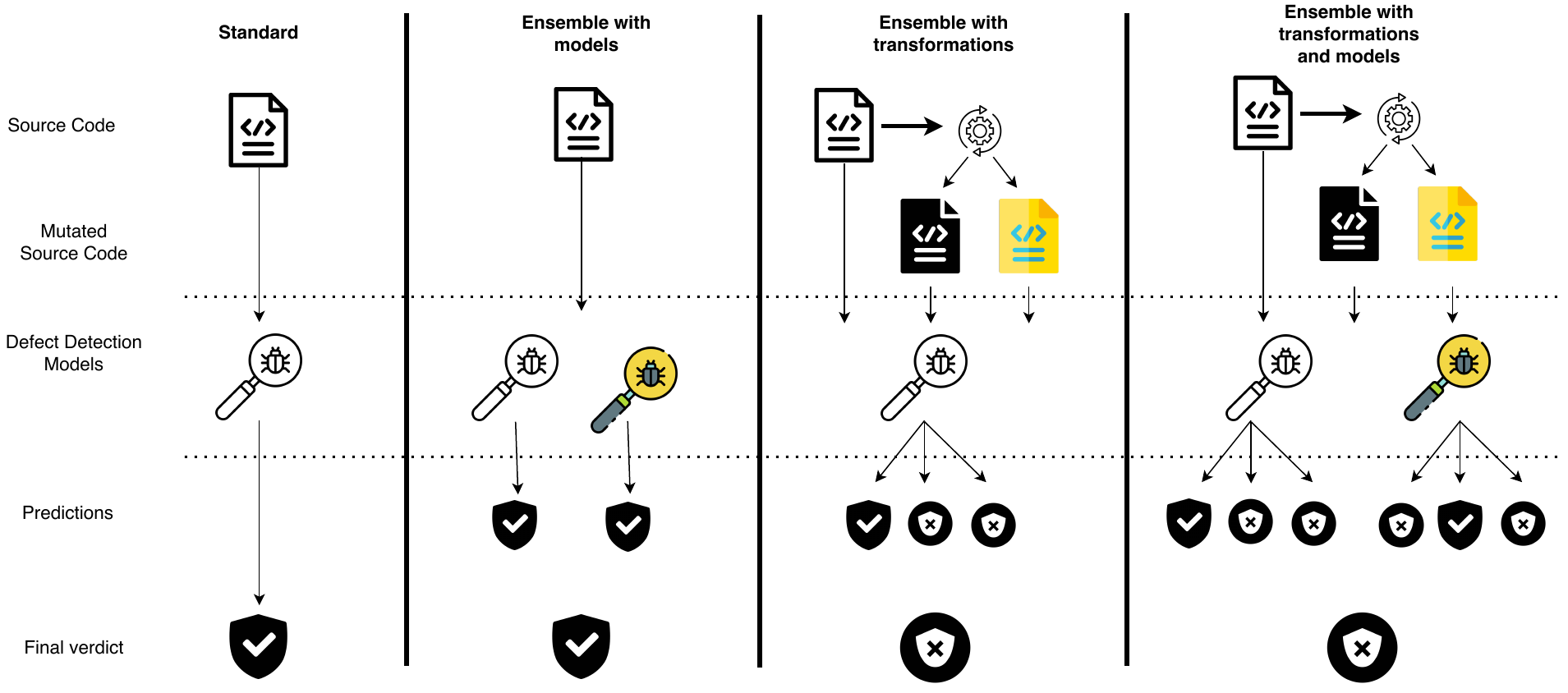}
\caption[]{Overview of the proposed approach for applying semantic-preserving transformation for testing defect detection models.}
\label{fig:overview}
\end{figure*}

To answer this question, we focus on Large Language Models (LLMs) as defect detection tools. 
First, we review existing literature that applied semantic-preserving transformations. 
Then we check whether they share resources to find transformations we can replicate and apply for our experiments.
Using LLMs as defect detection models and existing approaches for generating semantic-preserving transformations, we focus on combining the two parts and investigate the ability of semantic-preserving transformations to enhance and improve the defect detection models.
For this purpose, we use ensemble techniques, which make predictions under consideration of the original code snippet and mutated variants with semantic-preserving transformations.
Figure~\ref{fig:overview} outlines the current process of using LLMs for defect detection and our novel approach, using semantic-preserving transformations for mutating source code, and ensemble learning.
We apply the ensembles to the Devign dataset~\cite{zhou2019:devign}, a dataset for detecting vulnerability defects, which categorises functions as vulnerable and non-vulnerable.

We make the following contributions:
\begin{itemize}[noitemsep,topsep=0pt,parsep=0pt,partopsep=0pt]
    \item We collect 28 publications that applied semantic-preserving transformations. %
    \item We implement 39 semantic-preserving transformations shared by four publications. For each transformation, we check up to 20 transformed functions manually to verify their correctness.
    \item We evaluate the usability of semantic-preserving transformations for improving defect detection tools in an empirical study with three ensemble techniques and two LLMs using the transformations.
    \item We share our results and all transformations applied to the Devign dataset to enable a replication and verification of our results.\footnote{~\label{replication}
\url{https://figshare.com/s/f9b93d9d549f316b1e5d}.
}
\end{itemize}

\head{Main findings}
We have found \pubSemantic publications with \numTransforms semantic-preserving transformations. %
When implementing the transformations of four of these repositories, we encountered difficulties. Critically, 23 out of 39 transformations \textbf{did not pass our manual validation, as they changed semantics}. %
We are left with \appliedTransforms out of 39 transformations for our empirical study. 
While our investigated ensemble techniques were not able to improve the accuracy, we provide insights on implementation difficulties (Section~\ref{section:replication}).

\section{Related Work}
\label{section:rw}
\subsection{Data Augmentation for Software Engineering}
Data augmentation is used to increase the amount of data available for training or testing of machine learning models. 
Here we outline approaches for software engineering tasks.

\head{Sampling} The first set of approaches are concerned with the sampling of existing instances in the training dataset.
When treating sampling as a data augmentation approach, we focus on the oversampling of data (i.e., the addition of duplicates) rather than downsampling (i.e., removal of samples)~\cite{zagane2020:deep,aumpansub2021:detecting}. %

Sampling is in particular useful when dealing with imbalanced datasets, which frequently is the case for vulnerability detection tasks. 
For instance, datasets might only have 10\% of functions labeled as vulnerable~\cite{ferenc2019:challenging}.
When collecting programs from online sources, such as Github, the proportion of correct samples is much larger than defective programs.
Oversampling can balance the two classes~\cite{partenza2021:automatic}. %

In addition to random sampling, more complex methods have been applied, such as fuzzy-based oversampling~\cite{liu2020:deepbalance} or Synthetic Minority Over-Sampling (SMOTE)~\cite{jeon2021:autovas,chakraborty2020:deep:arxiv}.
SMOTE selects a sample from the under-represented group and one of its closest neighbors (from the same group). 
The new sample is created by interpolating between the two selected data samples~\cite{chawla2002:smote}.
For such an interpolation to work, one needs to represent programs and code snippets numerically. 
This has been done with vector embeddings~\cite{chakraborty2020:deep:arxiv,jeon2021:autovas} or based on source code metrics~\cite{kumar2018:application,kumar2019:method,pecorelli2019:role}

Regular oversampling is not able to generate new programs.
SMOTE can generate new data but represents programs as vector embeddings rather than source code.

\head{Neural Approaches} Neural approaches use LLMs to learn how to modify existing code snippets and create new snippets from scratch.
For example, Richter and Wehrheim~\cite{richter2022:learning} used LLMs to learn how to mutate tokens in a code snippet to generate more natural bugs.

Another approach has been followed by Zirak and Hemmati~\cite{zirak2022:improving}, who trained an LLM on a reverse program repair dataset such that they can use the LLM to generate faulty examples (e.g., learn how to create bugs rather than learning how to fix them).
For the bug detection and repair task, Allamanis et al.~\cite{allamanis2021:selfsupervised} trained two models, one to detect and repair bugs and a second one the learn how to best insert bugs to be used as training data.
An adaptation of this approach is introduced by Yasunaga and Jiang~\cite{yasunaga2021:breakitfixit}. They employed the iterative approach break-it-fix-it (BIFI) for data generation. BIFI uses a fixer, which learns to repair buggy code, and a breaker which learns to insert bugs in fixed code.
In an iterative process, both these models are used to improve one another (e.g., the fixer learns from the code ``broken'' by the breaker and vice versa).

While neural approaches are able to generate new source code samples, which look natural, they are usually applied to create datasets for program repair tasks (i.e., insert natural bugs for good code).
Thereby, the created programs aim at breaking the code and do not maintain the functionality of the original code snippets.

\head{Rule-based transformations}
Previous described approaches for data augmentation generate new, unseen samples but are not able to guarantee their behavior (e.g., neural approaches) or are not applicable to source code (e.g., SMOTE). 
One approach to address this shortcoming are rule-based transformations.
By carefully defining the modifications, one can ensure that the functionality of modified snippets is not altered (e.g., change for loops to while loops).
We note that one can also apply rule-based transformations to insert faults, but our focus here is the use of semantic-preserving transformations, transformation which can change the syntax of a code snippet but not the functionality.

Semantic-preserving transformations have been applied to a variety of software engineering tasks, including code clone detection~\cite{jain2021:contrastive} and code authorship disguise~\cite{quiring2019:misleading,liu2021:practical}.
The application ranges from the creation or extension of datasets to robustness evaluation and changes to pre-training procedures.
For example, Jing~\cite{jing2022:improvement} used transformations to increase the complexity of samples in a vulnerability detection dataset.
Other works tested the robustness of LLMs to small changes in the input. This means that a model should return the same prediction when faced with a code snippet and a transformed one with the same semantics~\cite{rabin2021:generalizability}.
Semantic-preserving transformations used to test the robustness can also be used to extend the training dataset and in turn improve a models' robustness~\cite{pour2021:searchbased,zhang2020:generating}.
Lastly, semantic-preserving transformation have been used to modify the loss function when training LLMs for software engineering tasks, one example being contrastive pre-training.
Contrastive pre-training~\cite{jain2021:contrastive,liu2023:contrabert,bui2021:selfsupervised} guides the training of LLMs such that semantically-identical code snippet are embedded close to one another.
Thereby, predictions by the model should be less dependent on syntax.

While rule-based semantic-preserving transformations have been applied to test the robustness of models, create or extend training datasets, they have not been applied to enhance and improve the testing itself. 
In this work, we are the first to use semantic-preserving transformations to extend the testing procedure, similar to metamorphic testing. 
We apply this to the defect detection task, in particular the detection of vulnerabilities.

\subsection{Ensemble Learning for Defect Detection}
Ensemble models combine predictions made by multiple models.
This combination is carried out to address limitations of machine learning models and can achieve higher accuracy than a single classifier~\cite{mienye2022:survey}.
In particular, multiple classification models can be used to complement one another, for example in situation where identifiers are able to detect different types of defects~\cite{panichella2014:crossproject,dinucci2017:dynamic}.
An ensemble of such classifiers could make for more robust defect detection models, for instance when dealing with vulnerabilities.

A survey on ensembles for defect prediction by Matloob et al.~\cite{matloob2021:software} found that the most frequent ensemble techniques to improve the performance of defect detection tools were random forests, boosting, and bagging. 
There was no mention of ensembles of different defect detection tools.

Di Nucci et al.~\cite{dinucci2017:dynamic} proposed ASCI (Adaptive Selection of ClassIfiers in bug prediction), an approach which instead of combining multiple classifiers into an ensemble, predicts the most suitable classifier to use. 
This approach was able to outperform a majority voting baseline.

In 2016, Petrić et al.~\cite{petric2016:building} stated that \textit{``Almost all previous work using ensemble techniques in defect prediction rely on the majority voting scheme for combining prediction outputs''}. 
Instead of applying a majority voting scheme, they used a stacking ensemble.
The first layer of a stacking approach contains predictions made by individual classification models. 
This is used as an input to the second layer, which learns to make the final prediction.

Further approaches for defect detection with ensemble learning used weighing~\cite{li2017:heterogeneous,dai2019:smash}, averaging prediction probabilities~\cite{laradji2014:software}, and stacked learning with different neural network architectures~\cite{wang2020:prennsem}.
Barbez et al.~\cite{barbez2019:machinelearning} used an ensemble of feature extraction tools to generate a comprehensive input for defect detection.
Other than predicting whether a function is vulnerable, Ding et al.~\cite{ding2022:velvet} used ensemble learning to detect vulnerabilities on a statement level.

\begin{table}
\vspace{2mm}
\caption{Semantic-preserving transformations - Part 1.}
\label{table:transforms1}
\newlength{\firstcol}
\setlength{\firstcol}{5em}
\begin{adjustbox}{max width=\columnwidth}
\rowcolors{1}{}{Gray2}
\begin{tabular}{lllp{2.5cm}}
\toprule
Types & Transformation & \# & Reference \\ \midrule
\cellcolor{white}  & Replace variable name & 25 & \cite{liu2021:practical, pour2021:searchbased, tian2023:adversarial, yefet2020:adversarial, bielik2020:adversarial, applis2021:assessing, dong2023:boosting, wei2021:cocofuzzing, liu2023:contrabert, jain2021:contrastive, dong2022:enhancing, shi2022:enhancing, srikant2021:generating, zhang2020:generating, quiring2019:misleading, chakraborty2022:natgen, yang2022:natural, rabin2021:generalizability, li2022:ropgen, applis2023:searching, bui2021:selfsupervised, ramakrishnan2022:semantic, yang2023:stealthy, springer2021:strata, allamanis2021:selfsupervised} \\
\cellcolor{white}  & Replace argument/parameter & 8 & \cite{pour2021:searchbased, dong2023:boosting, jain2021:contrastive, dong2022:enhancing, srikant2021:generating, risse2023:limits, applis2023:searching, ramakrishnan2022:semantic} \\
\cellcolor{white}  & Replace function name & 8 & \cite{liu2021:practical, pour2021:searchbased, applis2021:assessing, dong2023:boosting, liu2023:contrabert, dong2022:enhancing, risse2023:limits, quiring2019:misleading} \\
\cellcolor{white}  & Object field renaming & 3 & \cite{bielik2020:adversarial, srikant2021:generating, ramakrishnan2022:semantic} \\
\cellcolor{white} & Split/aggregate declarations & 3 & \cite{liu2021:practical, tian2021:generating, li2022:ropgen} \\
\cellcolor{white}  & Split/aggregate multi-variable assignment & 3 & \cite{tian2021:generating, chakraborty2022:natgen, li2022:ropgen} \\
\cellcolor{white}  & API renaming & 3 & \cite{pour2021:searchbased, dong2023:boosting, dong2022:enhancing} \\
\cellcolor{white}  & Swap operands & 3 & \cite{liu2021:practical, chakraborty2022:natgen, allamanis2021:selfsupervised} \\
\cellcolor{white}  & Undo or introduce type alias & 3 & \cite{liu2021:practical, quiring2019:misleading, li2022:ropgen} \\
\cellcolor{white}  & Split/aggregate assignment initialization & 2 & \cite{tian2021:generating, li2022:ropgen} \\
\cellcolor{white}  & Prefix / suffix operator swap & 2 & \cite{tian2021:generating, li2022:ropgen} \\
\cellcolor{white}  & Separate/attach elaborated type declaration & 1 & \cite{liu2021:practical} \\
\cellcolor{white}  & Replace a class & 1 & \cite{applis2021:assessing} \\
\cellcolor{white}  & Property assignment renaming & 1 & \cite{bielik2020:adversarial} \\
\cellcolor{white}  & Use namespace & 1 & \cite{li2022:ropgen} \\
\cellcolor{white}  & Use macros & 1 & \cite{li2022:ropgen} \\
\cellcolor{white}  & Use alternative tokens & 1 & \cite{liu2021:practical} \\
\cellcolor{white}  & Use converse-negative expressions & 1 & \cite{liu2021:practical} \\
\cellcolor{white}  & Use equivalent computations & 1 & \cite{liu2021:practical} \\
\cellcolor{white}  & Numerical calculation transformation & 1 & \cite{tian2023:adversarial} \\
\cellcolor{white}  & Self-increasing/decreasing unfolding & 1 & \cite{tian2021:generating} \\
\cellcolor{white}  & Ternary expressions & 1 & \cite{bielik2020:adversarial} \\
\cellcolor{white}  & Array access & 1 & \cite{bielik2020:adversarial} \\
\multirow{-24}{\firstcol}{\cellcolor{white}Trivial} & Array indexing/pointer & 1 & \cite{li2022:ropgen} \\ \midrule

\cellcolor{white} & Add neutral element & 5 & \cite{applis2021:assessing, dong2023:boosting, wei2021:cocofuzzing, dong2022:enhancing, applis2023:searching} \\ 
\cellcolor{white}  & If enhancement & 3 & \cite{pour2021:searchbased, dong2023:boosting, dong2022:enhancing} \\
\cellcolor{white}  & Convert int literals into expressions, vice versa & 2 & \cite{liu2021:practical, jain2021:contrastive} \\
\cellcolor{white}  & Convert between bool literals and int literals & 2 & \cite{liu2021:practical, quiring2019:misleading} \\
\cellcolor{white}  & Replace bool literals & 2 & \cite{srikant2021:generating, ramakrishnan2022:semantic} \\
\cellcolor{white}  & Boolean exchange & 2 & \cite{shi2022:enhancing, rabin2021:generalizability} \\
\cellcolor{white}  & For loop enhancement & 2 & \cite{dong2023:boosting, dong2022:enhancing} \\
\cellcolor{white}  & Convert integers into hexadecimal numbers & 1 & \cite{liu2021:practical} \\
\cellcolor{white}  & Promote data types & 1 & \cite{quiring2019:misleading} \\
\cellcolor{white}  & Convert char literals into ASCII values & 1 & \cite{liu2021:practical} \\
\cellcolor{white}  & Convert string literals to char arrays & 2 & \cite{liu2021:practical, quiring2019:misleading} \\
\cellcolor{white}  & Convert array to vector & 1 & \cite{quiring2019:misleading} \\
\cellcolor{white}  & Declaration of loop variables & 1 & \cite{quiring2019:misleading} \\
\cellcolor{white}  & Type upconversion & 1 & \cite{jain2021:contrastive} \\
\cellcolor{white}  & Use cast expressions & 1 & \cite{liu2021:practical} \\
\multirow{-16}{\firstcol}{\cellcolor{white} Data and Declaration} & Use typeid expression & 1 & \cite{liu2021:practical} \\ \midrule

\cellcolor{white} & Input API transformation & 3 & \cite{tian2021:generating, quiring2019:misleading, li2022:ropgen} \\
\cellcolor{white} & Output API transformation & 3 & \cite{tian2021:generating, quiring2019:misleading, li2022:ropgen} \\
\cellcolor{white} & Input interface transformer & 2 & \cite{quiring2019:misleading, li2022:ropgen} \\
\cellcolor{white} & Output interface transformer & 2 & \cite{quiring2019:misleading, li2022:ropgen} \\
\multirow{-5}{\firstcol}{\cellcolor{white}API} & Sync-with-stdio transformer & 2 & \cite{quiring2019:misleading, li2022:ropgen} \\ \bottomrule
\end{tabular}
\end{adjustbox}
\end{table}

\begin{table}
\vspace{2mm}
\caption{Semantic-preserving transformations - Part 2.}
\label{table:transforms2}
\setlength{\firstcol}{5em}
\begin{adjustbox}{max width=\columnwidth}
\rowcolors{1}{}{Gray2}
\begin{tabular}{lllp{3cm}}
\toprule
Types & Transformation & \# & Reference \\ \midrule
\cellcolor{white}  & Convert for-statement to while-statement & 9 & \cite{liu2021:practical, pour2021:searchbased, tian2023:adversarial, tian2021:generating, quiring2019:misleading, chakraborty2022:natgen, rabin2021:generalizability, li2022:ropgen, bui2021:selfsupervised} \\
\cellcolor{white}  & Convert while-statement to for-statement & 9 & \cite{liu2021:practical, pour2021:searchbased, tian2023:adversarial, tian2021:generating, quiring2019:misleading, chakraborty2022:natgen, rabin2021:generalizability, li2022:ropgen, bui2021:selfsupervised} \\
\cellcolor{white}  & Convert switch-case to if-else & 6 & \cite{liu2021:practical, shi2022:enhancing, tian2021:generating, rabin2021:generalizability, li2022:ropgen, bui2021:selfsupervised} \\
\cellcolor{white}  & Reorder statements & 5 & \cite{liu2023:contrabert, shi2022:enhancing, rabin2021:generalizability, li2022:ropgen, bui2021:selfsupervised} \\
\cellcolor{white}  & Convert if-else to switch-case & 4 & \cite{liu2021:practical, tian2021:generating, li2022:ropgen, bui2021:selfsupervised} \\
\cellcolor{white}  & Split conditions of if-statements & 4 & \cite{liu2021:practical, tian2021:generating, quiring2019:misleading, li2022:ropgen} \\
\cellcolor{white}  & Swap if-else bodies & 3 & \cite{liu2021:practical, chakraborty2022:natgen, allamanis2021:selfsupervised} \\
\cellcolor{white}  & Convert ternary to if & 2 & \cite{tian2021:generating, chakraborty2022:natgen} \\
\cellcolor{white}  & If-true & 2 & \cite{applis2021:assessing, applis2023:searching} \\
\cellcolor{white}  & Lambda-identity & 2 & \cite{applis2021:assessing, applis2023:searching} \\
\cellcolor{white}  & Combine if statement & 2 & \cite{tian2021:generating, li2022:ropgen} \\
\cellcolor{white}  & Convert if-else to conditional expression & 1 & \cite{liu2021:practical} \\
\cellcolor{white}  & Convert conditional expression to if-else & 1 & \cite{liu2021:practical} \\
\cellcolor{white}  & Convert if-else(if) to if-if & 1 & \cite{tian2023:adversarial} \\
\cellcolor{white}  & Unroll while loop & 1 & \cite{ramakrishnan2022:semantic} \\
\cellcolor{white}  & Insert multiple loops & 1 & \cite{jing2022:improvement} \\
\cellcolor{white}  & Wrap Try Catch & 1 & \cite{ramakrishnan2022:semantic} \\
\multirow{-18}{\firstcol}{\cellcolor{white}Control flow} & Delegation-method & 1 & \cite{applis2021:assessing} \\ \midrule

\cellcolor{white}  & Add function arguments & 5 & \cite{liu2021:practical, pour2021:searchbased, applis2021:assessing, dong2023:boosting, dong2022:enhancing} \\
\cellcolor{white}  & Function creation & 3 & \cite{risse2023:limits, quiring2019:misleading, li2022:ropgen} \\
\cellcolor{white} & Reorder function arguments & 2 & \cite{liu2021:practical, risse2023:limits} \\
\cellcolor{white}  & Add input checking for function parameters & 2 & \cite{dong2023:boosting, dong2022:enhancing} \\ 
\cellcolor{white}  & Merge function arguments & 1 & \cite{liu2021:practical} \\
\cellcolor{white}  & Convert statements into functions & 1 & \cite{liu2021:practical} \\
\cellcolor{white}  & Convert binary expressions into functions & 1 & \cite{liu2021:practical} \\
\cellcolor{white}  & Merge functions & 1 & \cite{liu2021:practical} \\
\cellcolor{white}  & Hide API calls & 1 & \cite{liu2021:practical} \\
\multirow{-9}{\firstcol}{\cellcolor{white}Function} & Merge function arguments & 1 & \cite{liu2021:practical} \\ \midrule

\cellcolor{white}  & Add dead code & 12 & \cite{dong2023:boosting, wei2021:cocofuzzing, liu2023:contrabert, jain2021:contrastive, dong2022:enhancing, shi2022:enhancing, srikant2021:generating, jing2022:improvement, risse2023:limits, chakraborty2022:natgen, bui2021:selfsupervised, ramakrishnan2022:semantic} \\
\cellcolor{white}  & Add unused variable & 10 & \cite{pour2021:searchbased, yefet2020:adversarial, bielik2020:adversarial, applis2021:assessing, dong2023:boosting, wei2021:cocofuzzing, dong2022:enhancing, risse2023:limits, rabin2021:generalizability, applis2023:searching} \\
\cellcolor{white} & Add temp variables & 5 & \cite{liu2021:practical, tian2023:adversarial, jing2022:improvement, quiring2019:misleading, li2022:ropgen} \\
\cellcolor{white}  & Add print statement & 5 & \cite{pour2021:searchbased, dong2023:boosting, dong2022:enhancing, srikant2021:generating, ramakrishnan2022:semantic} \\
\cellcolor{white}  & Add, remove, move comments & 5 & \cite{applis2021:assessing, liu2023:contrabert, jain2021:contrastive, risse2023:limits, allamanis2021:selfsupervised} \\
\cellcolor{white}  & Return optimal & 3 & \cite{pour2021:searchbased, dong2023:boosting, dong2022:enhancing} \\
\cellcolor{white}  & Duplication & 3 & \cite{dong2023:boosting, wei2021:cocofuzzing, dong2022:enhancing} \\
\cellcolor{white}  & Add libraries/includes & 3 & \cite{liu2021:practical, quiring2019:misleading, li2022:ropgen} \\
\cellcolor{white}  & Add global declarations & 3 & \cite{liu2021:practical, quiring2019:misleading, li2022:ropgen} \\
\cellcolor{white}  & Remove dead code & 3 & \cite{liu2021:practical, jain2021:contrastive, quiring2019:misleading} \\
\cellcolor{white}  & Add return statement & 2 & \cite{quiring2019:misleading, li2022:ropgen} \\
\cellcolor{white}  & Add redundant operands & 2 & \cite{liu2021:practical, wei2021:cocofuzzing} \\
\cellcolor{white}  & Add type alias & 2 & \cite{liu2021:practical, quiring2019:misleading} \\
\cellcolor{white}  & Add/remove compound statement & 1 & \cite{quiring2019:misleading} \\
\cellcolor{white}  & Add unused object expression & 1 & \cite{bielik2020:adversarial} \\
\cellcolor{white}  & Add function declarations in classes & 1 & \cite{liu2021:practical} \\
\multirow{-17}{\firstcol}{\cellcolor{white}Dead/bogus code} & Add void function & 1 & \cite{risse2023:limits} \\ \midrule

\cellcolor{white} & Add whitespace & 2 & \cite{applis2021:assessing, risse2023:limits} \\
\cellcolor{white} & Remove whitespace & 1 & \cite{applis2021:assessing} \\
\cellcolor{white} & Reformatting & 1 & \cite{jain2021:contrastive} \\
\cellcolor{white} & Beautification & 1 & \cite{jain2021:contrastive} \\
\multirow{-5}{\firstcol}{\cellcolor{white}Formatting} & Compression & 1 & \cite{jain2021:contrastive} \\ \bottomrule
\end{tabular}
\end{adjustbox}
\end{table}

\section{Transformations}
\label{sec:transformations}
In this section, we present an overview of semantic-preserving transformations that have been applied in existing works.
For this, we performed and exploratory literature search in addition to backward snowballing. 
We found \pubSemantic publications that have applied a total of \numTransforms different transformations.
Table~\ref{table:transforms1} and Table~\ref{table:transforms2} list the \numTransforms transformations.
A description of each transformation can be found in our online appendix.\footref{replication}

We further divide the transformations in six categories, inspired by Quiring et al.~\cite{quiring2019:misleading} and Liu et al.~\cite{liu2021:practical}: API, Formatting, Control Flow, Function, Data and Declaration, Dead/bogus code, Trivial.

\textbf{API} transformations are concerned with API use, mainly for input and output processing. 
\textbf{Formatting} transformations do not change any of the tokens of code snippets, but rather adjust spacing.
\textbf{Control Flow} transformations describe modifications to the snippets control flow, such as modification of if-statements or the processing order of loop operations.
These are often times bi-directional, such as the conversion of for loops to while loops and vice-versa.
\textbf{Function} transformations are concerned with transformations that address the modification of functions (e.g., function parameters) or the creation of new functions (e.g., the conversion of statements to functions or the merging of two functions).
\textbf{Data and Declaration} addresses transformations such as the declaration of variables (e.g., defining multiple variables in a single statement) and the conversion of different data types (e.g., casting integers to floats).
\textbf{Dead/bogus Code} describes the addition of useless statements to the code, which do not change the execution, such as duplicating assignments or modifying comments.
Unused code can describe a multitude of constructs, such as if, for, if else, switch or while statements~\cite{dong2023:boosting,wei2021:cocofuzzing,dong2022:enhancing}.
\textbf{Trivial} transformations include all other, small modifications to the code, such as renaming (functions, variable names, arguments), rewriting of operands (i++ is changed to i+=1), or the splitting/aggregation of statements.
The most frequently applied transformation belongs to this category, i.e., the renaming of variable names.
Variable renaming has been performed by 25 out of \pubSemantic publications, following different strategies, such as the use of synonyms~\cite{pour2021:searchbased}, setting the variable name to $VAR_i$~\cite{chakraborty2022:natgen}, adversarial selection~\cite{yang2022:natural} and random generation~\cite{wei2021:cocofuzzing}.

\begin{table*}[!htb]
\vspace{2mm}
\caption{Overview of publications with shared source code for transformations and the respective programming languages (PLs), including the number of implemented transformations (\#).}
\label{table:shared}
\begin{adjustbox}{max width=\textwidth}
\rowcolors{1}{}{Gray2}
\begin{tabular}{lp{5cm}lp{3cm}p{9cm}}
\toprule
Author, year & Topic & \#  & PLs & Link \\ \midrule
Zhen et al.~\cite{li2022:ropgen}, 2022 & Authorship attribution & 26 & C/C++, Java & \url{https://github.com/RoPGen/RoPGen} \\
Quiring et al.~\cite{quiring2019:misleading}, 2019 & Authorship attribution & 24 & C/C++ & \url{https://github.com/EQuiw/code-imitator} \\
Dong et al.~\cite{dongMIXCODEEnhancingCode2023}, 2023 & Code classification, Defect detection & 14 & Java, Python & \url{https://github.com/zemingd/Mixup4Code} \\
Applis et al.~\cite{applis2021:assessing}, 2021 & Code summarization & 12 & Java & \url{https://github.com/ciselab/Lampion/tree/main/Transformers} \\
Pour et al.~\cite{pour2021:searchbased}, 2021 & Method name prediction, code captioning, code search, documentation generation & 11 & Java & \url{https://github.com/MaryamVP/Guided-Mutation-ICST-2021} \\
Jain et al.~\cite{jain2021:contrastive}, 2021 & Code clone detection, type inference, Extreme code summarization & 10 & JavaScript, TypeScript & \url{https://github.com/parasj/contracode} \\
Risse and Boehme~\cite{risse2023:limits}, 2023 & Vulnerability detection & 9 & C/C++ & \url{https://github.com/niklasrisse/LimitsOfML4Vuln} \\
Ramakrishnan et al.~\cite{ramakrishnan2022:semantic}, 2022 & Method name prediction & 8 & Java, Python & \url{https://github.com/jjhenkel/averloc} \\
Chakraborty et al.~\cite{chakraborty2022:natgen}, 2022 & Text to code generation, code translation, bug fixing & 8 & Java, Python, C, C\#, Go, JavaScript, Ruby, PHP & \url{https://github.com/saikat107/NatGen} \\
Rabin et al.~\cite{rabin2021:generalizability}, 2021 & Method name prediction & 7 & Java & \url{https://github.com/mdrafiqulrabin/tnpa-generalizability/} \\
Bielik et al.~\cite{bielik2020:adversarial}, 2020 & type prediction & 7 & JavaScript, TypeScript & \url{https://github.com/eth-sri/robust-code} \\
Wei et al.~\cite{wei2021:cocofuzzing}, 2021 & NL summary, Method name prediction & 6 & Java & \url{https://zenodo.org/record/4000441#.ZEeX2-xByX0} \\
Srikant et al.~\cite{srikant2021:generating}, 2021 & Method name prediction & 6 & Java, Python & \url{https://github.com/ALFA-group/adversarial-code-generation} \\
Liu et al.~\cite{liu2023:contrabert}, 2023 & Clone detection, defect detection, code translation, code search & 5 & Java, Python, Ruby, Go, PHP, JavaScript & \url{https://github.com/shangqing-liu/ContraBERT} \\
Allamanis et al.~\cite{allamanis2021:selfsupervised}, 2021 & Bug detection, program repair & 4 & Python & \url{https://github.com/microsoft/neurips21-self-supervised-bug-detection-and-repair} \\
Yefet et al.~\cite{yefet2020:adversarial}, 2020 & Code classification, Bug detection & 2 & Java, C\# & \url{https://github.com/tech-srl/adversarial-examples} \\
Zhang et al.~\cite{zhang2020:generating}, 2020 & Source code classification & 1 & C/C++ & \url{https://github.com/SEKE-Adversary/MHM} \\
Yang et al.~\cite{yang2022:natural}, 2022 & Vulnerability prediction, Clone detection, Authorship attribution & 1 & C, Python, Java & \url{https://github.com/soarsmu/attack-pretrain-models-of-code} \\
Yang et al.~\cite{yang2023:stealthy}, 2023 & Code summarization NL, method name prediction & 1 & Python & \url{https://figshare.com/articles/dataset/ICSE-23-Replication_7z/20766577/1} \\
\bottomrule
\end{tabular}
\end{adjustbox}
\end{table*}

\subsection{Publications with Shared Artifacts}
In addition to collecting publications with semantic-preserving transformation, we are particularly interested in the ones that share their implementation.
A shared implementation allows for a reproduction of the results and application of transformations without incurring large implementation overheads and the risk of errors or different implementation choices.
Among the \pubSemantic publications that applied semantic-preserving transformations, we found \pubRelevant which share code, such that one can apply their transformations.
Table~\ref{table:shared} provides details on each of the \pubRelevant publications. 
For each publication, we describe the task to which they have been applied to (e.g., code search, clone detection), the number of implemented transformations (according to Table~\ref{table:transforms1} and Table~\ref{table:transforms2}), the respective programming languages as well as a link to their resources.

From Table~\ref{table:shared}, we can observe that the shared transformation have been applied to various tasks, ranging from defects (vulnerability detection, program repair), code translation to code clone detection and authorship attribution.
Moreover, there are several publications that applied a single transformation, i.e., variable renaming.

In total, shared transformations have been applied to 9 programming languages. 
The most frequently used programming language is Java, with 12 out of \pubRelevant publications. 
The next popular programming language is Python (8 publications) followed by C/C++ (6 publications) and JavaScript (4 publications).
In 9 out of \pubRelevant cases, transformations have been applied to a single programming language, the remaining 10 consider at least two and at most 8.

\section{Empirical Study Design}

\subsection{Research Questions}
\noindent
We set out to answer two research questions in our study:

\textbf{RQ1. What is the impact of semantic-preserving transformation operators on the predictions made by defect detection tools?}

\noindent
In the first RQ, we investigate whether semantic-preserving transformations lead to changes in the predictions of defect detection tools.
In particular, we are interested in vulnerability defects in code snippets.
This question is important to verify the viability of our approach for improving vulnerability detection (e.g., if the transformed code does not change predictions, the information cannot be used for improvement).
Moreover, we investigate how often each transformation can be applied to the functions of the considered dataset (Section~\ref{section:dataset}). 
These insights support our second research question:

\textbf{RQ2. To what extent can transformation be used to improve defect detection tools?}

\noindent
To answer this question, we augment the defect detection tools, normally applied to a single function, to incorporate predictions performed on the transformations.
This resembles an ensemble strategy (e.g., consider the majority prediction of the original and $x$ transformed functions).
More details on the implemented ensemble approaches are provided in Section~\ref{section:ensemble}.

\subsection{Dataset}
\label{section:dataset}
We use the Devign~\cite{zhou2019:devign} vulnerability detection dataset to evaluate semantic preserving transformations, which is part of the ML benchmark collection CodeXGLUE~\cite{lu2021:codexglue}.
The Devign dataset~\cite{zhou2019:devign} includes $27,318$ C/C++ functions from open-source projects, labeled vulnerable or non-vulnerable (54\% of the functions are labeled non-vulnerable).\footnote{\url{https://drive.google.com/file/d/1x6hoF7G-tSYxg8AFybggypLZgMGDNHfF}}
Here, the label $1$ indicates a vulnerable and the label $0$ a non-vulnerable function.
The data is split as follows: 80\% training, 10\% testing, 10\% validation.\footnote{~\url{https://github.com/microsoft/CodeXGLUE/tree/main/Code-Code/Defect-detection}}
To determine the quality of defect detection tools applied to the Devign dataset, we measure their accuracy (i.e., the proportion of correctly labeled functions), in accordance with CodeXGLUE~\cite{lu2021:codexglue}.

\subsection{Models}
\begin{table*}
\caption{Vulnerability detection models and their shared resources. Accuracy is given in accordance with CodeXGLUE.}
\label{table:models}
\begin{adjustbox}{max width=\textwidth}
\rowcolors{1}{}{Gray2}
\begin{tabular}{lccccl}
\toprule
Model & Accuracy & Fine-tuned & Pre-trained & Code & URLs \\ \midrule
UniXcoder-nine-MLP~\cite{guo2022:unixcoder} & 69.29 &  & \checkmark & \checkmark & \url{https://github.com/microsoft/CodeBERT/tree/master/UniXcoder} \\
CoTexT~\cite{phan2021:cotext} & 66.62 &  & \checkmark &  & \url{https://huggingface.co/razent/cotext-1-ccg} \\
C-BERT~\cite{buratti2020:exploring} & 65.45 &  &  &  &  \\
A-BERT & 65.37 &  &  &  &  \\
RefactorBERT & 65.08 &  &  &  &  \\
VulBERTa-MLP~\cite{hanif2022:vulberta} & 64.75 & \checkmark & \checkmark & \checkmark & \url{https://github.com/ICL-ml4csec/VulBERTa} \\
VulBERTa-CNN~\cite{hanif2022:vulberta} & 64.42 & \checkmark & \checkmark & \checkmark & \url{https://github.com/ICL-ml4csec/VulBERTa} \\
ContraBERT\_C~\cite{liu2023:contrabert} & 64.17 &  & \checkmark & \checkmark & \url{https://github.com/shangqing-liu/ContraBERT} \\
ContraBERT\_G~\cite{liu2023:contrabert} & 63.32 &  & \checkmark & \checkmark & \url{https://github.com/shangqing-liu/ContraBERT}   \\
PLBART~\cite{ahmad2021:unified} & 63.18 & \checkmark & \checkmark & \checkmark & \url{https://github.com/wasiahmad/PLBART} \\
code2vec~\cite{alon2019:code2vec,coimbra2021:using} & 62.48 &  & \checkmark & \checkmark & \url{https://github.com/dcoimbra/dx2021}, \url{https://github.com/tech-srl/code2vec} \\
CodeBERT~\cite{feng2020:codebert} & 62.08 &  & \checkmark & \checkmark & \url{https://github.com/microsoft/CodeXGLUE/tree/main/Code-Code/Defect-detection} \\
\bottomrule
\end{tabular}
\end{adjustbox}

\end{table*}

Table~\ref{table:models} provides an overview of all the LLMs applied to the defect detection task on CodeXGLUE~\cite{lu2021:codexglue}. 
We list LLMs according to their accuracy on the Devign test set and provide details on the type of artifacts shared.
For each LLM, source code, pre-trained models and/or fine-tuned variants can be shared.
Here, fine-tuned indicates whether a model has been fine-tuned on the Devign training set and shared.
We observe that there are only two LLMs for which fine-tuned versions are shared: VulBERTa and PLBART.
Therefore, we use VulBERTa-MLP (the better performing variant of both VulBERTa versions) and PLBART to investigate semantic preserving transformations without incurring training costs.

We have run these models on one NVIDIA Tesla V100 SXM3 32 GB GPU.

\subsection{Semantic Preserving Transformations}
\changed{
To determine semantic-preserving transformations to implement from existing works (Table~\ref{table:transforms1} \&~\ref{table:transforms2}), we first select relevant publications to replicate according to three criteria:
\begin{enumerate}
    \item Implementation is available.
    \item Multiple transformations are implemented.
    \item Transformations are applicable to C/C++.
\end{enumerate}
The first stage of filtering leaves us with \pubRelevant publications, which can be seen in Table~\ref{table:shared}.
Seven of these can be applied to C/C++ code, which is required given the Devign dataset (Section~\ref{section:dataset}).
Next, we filter publications that implemented only a single transformation. 
We decided for this filtering to achieve a good trade-off between the effort required to implement shared repositories and the total number of transformations.
This leaves us with repositories from four publications: NatGen~\cite{chakraborty2022:natgen}, CodeImitator~\cite{quiring2019:misleading}, LimitsOfML4Vuln~\cite{risse2023:limits}, RoPGen~\cite{li2022:ropgen}.
These publications provide a total of 39 transformations, for which we obtained implementations from the shared repositories. 
Lastly, we perform another filtering stage to determine which of these 39 transformations to keep:
\begin{enumerate}
    \setcounter{enumi}{3}
    \item A manual check confirms that the transformations are indeed semantically-preserving. 
\end{enumerate}
For this purpose, we apply the transformations to functions of the Devign dataset and manually checked the correctness of up to 20 of them (i.e., we checked at least 20 transformed functions to confirm correctness but might terminate earlier when we have encountered incorrect transformations). 
We perform this check for each the the 39 transformations.
}
If any of those \textbf{did change the semantics of the function}, we excluded the transformation for further experiments, which was the case for 23 of them.
We are left with \appliedTransforms transformations without encountered errors. %
These constitute all transformation that we could implement from the four existing publications without encountering errors (i.e., the other 23 transformations changed semantics contrary to their claim).
We provide an overview of the investigated transformed functions and our verdict, whether they are semantic-preserving, in our online appendix.\footref{replication}

\subsection{Ensemble Strategies}
\label{section:ensemble}
We consider three types of ensembles (see Figure~\ref{fig:overview}):
\begin{enumerate}
    \item Ensemble with a single model and multiple functions;
    \item Ensemble with multiple models and a single function;
    \item Ensemble with multiple models and multiple functions.
\end{enumerate}

We follow popular ensemble approaches, such as majority voting and weighting~\cite{yang2023:survey,mienye2022:survey}.
\textbf{Majority Voting} makes predictions based on the majority prediction from a collection of prediction (e.g., if 7 out of 10 predictions are positive, the final verdict is positive).
We consider two variants of majority voting, depending on the strategy to break ties: selecting the label $0$ or the label $1$.

The second strategy we employ is \textbf{averaging}.
Rather than considering predictions as binary values (either 0 or 1), averaging uses the predicted probabilities (ranging from 0 to 1), and averages them among all available predictions.
A potential advantage of averaging is that it takes ``certainty'' in account, e.g., a value of 0.52 receives the prediction $1$ while being close to the decision boundary (0.5) is treated identical to 0.98.

Lastly, we employ a \textbf{weighting} strategy. 
Weighting can be compared to stacking, which builds a prediction model on top of the available predictions.
While majority voting and averaging give an equal weight to all available predictions, weighting assigns specific weights to each:

\footnotesize
\begin{equation}\label{eqn:weighted}
    pred = w_{17}*predVulberta + w_{18}*predPlbart +\sum_{n=1}^{16} w_n*predT_n
\end{equation}
\normalsize
Here, $predT_n$ shows the predictions made on a transformed function, using transformation $Tn$. $predVulberta$ and $predPlbart$ are the predictions performed on the original function by the two LLMs.
The final prediction ($pred$) requires a total of 18 weights ($w_1, ... , w_{18}$): original prediction from VulBERTa, prediction from PLBART, 16 transformations.\footnote{In our implementation, $w_1$ and $w_2$ are used for the original functions and the later 16 for the transformations.}

We follow two strategies to handle weighting: based on labels, based on probabilities.
When treating the predictions based on \textbf{labels}, we convert the label $0$ to -1 while a label of 1 remains unchanged. 
\changed{This allows us to sum up predictions and better aggregate the predictions obtained on transformed functions.
For instance, if $T_1$ can be applied to a function four times, for which converted labels are [-1,-1,-1,1], $predT_1$ is assigned a score equal to the sum of predicted labels: $predT_1 = \sum [-1,-1,-1,1] = -2$.
Without converting labels, we would receive $predT_n = \sum [0,0,0,1] = 1$.
}
Similarly, we adjust predictions based on \textbf{probabilities} in a range of $-0.5$ to $0.5$.
We do this by subtracting 0.5 from the probability of predicting the label $1$.
For example, if the probabilities to predict each of the two labels is $[0.2, 0.8]$, we treat it as $0.3$ ($0.8-0.5$).

To find the weights, we treat the Equation~\ref{eqn:weighted} as an optimization problem to find weights which maximize accuracy on the validation set.
We use optimizers from SciPy~\cite{virtanen2020:scipy} to obtain the weights.\footnote{We used the following eight optimizers: Nelder-Mead,Powell,CG,BFGS,L-BFGS-B,TNC,COBYLA,SLSQP}

\subsection{Threats to Validity}
Here we address threats with regards to our implementation and analysis (threats to internal validity) and the generalizability of results (threats to external validity).

To reduce threats to internal validity, we use existing repositories and implementations for our experiments.
In particular, every semantic-preserving transformation we applied is based on the implementation shared by published work.
\changed{To verify whether the transformations indeed preserve semantics, we carried out a manual check of up to 20 transformed functions. While we were able to exclude 23 incorrect transformations in this way, there might still be undiscovered faults which more extensive checks might be able to expose.}
Moreover, the two investigated LLM have already been fine-tuned, removing the risk of any mistakes in the training procedure. 
\changed{Given the benchmarking provided by CodeXGLUE and the fact that the Devign dataset is balanced, we opted for using accuracy to determine the performance of LLMs.}
However, we note that other metrics, such as F-score, could be considered as well.
Lastly, we share our results online to allow for replicability.

To reduce threats to external validity, we applied a wide range of semantic-preserving transformations, three ensemble strategies and two pre-trained LLMs. 
The performance is evaluated on the Devign dataset, a defect detection datasets with real-world bugs. 
An extension of experimental artifacts could further improve the generalizability of results.

\section{Results \& Discussion}

\subsection{RQ1: Semantic-Preserving Transformations}
\begin{figure}
\centering
\includegraphics[width=.5\columnwidth]{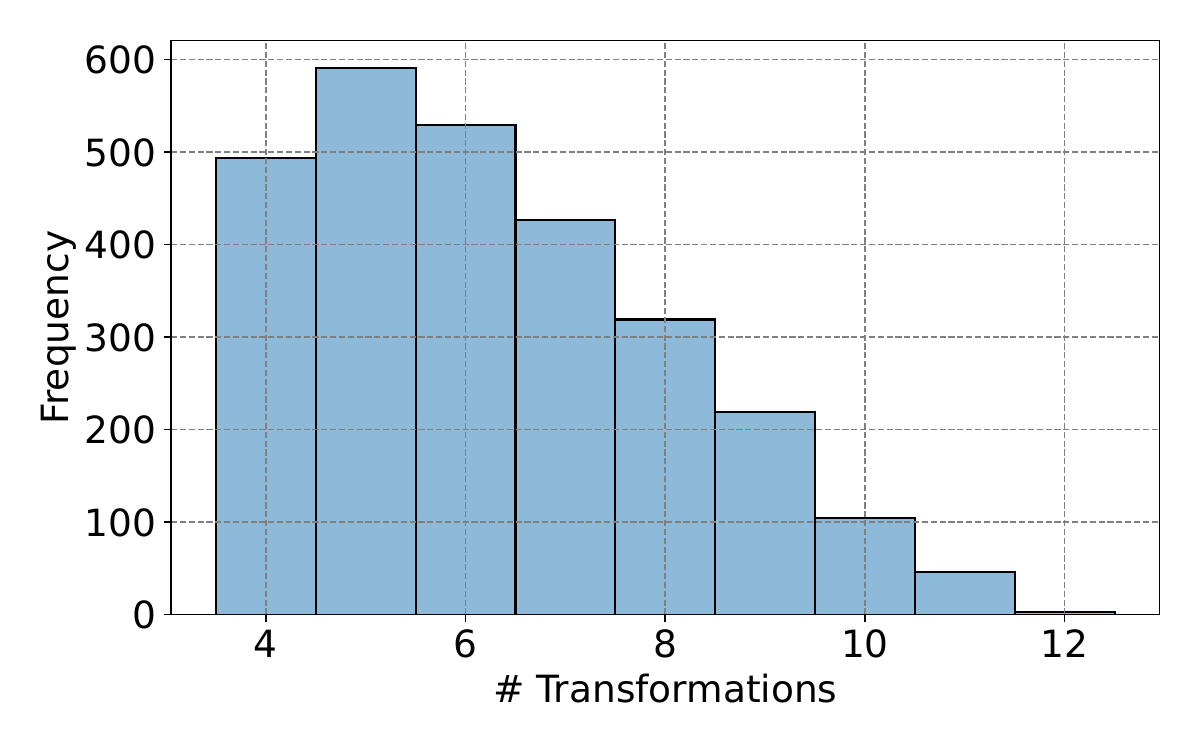}
\caption[]{RQ1: Number of transformations per function in the Devign test set.}
\label{fig:histogram}
\end{figure}

\begin{figure*}
\includegraphics[width=.9\textwidth]{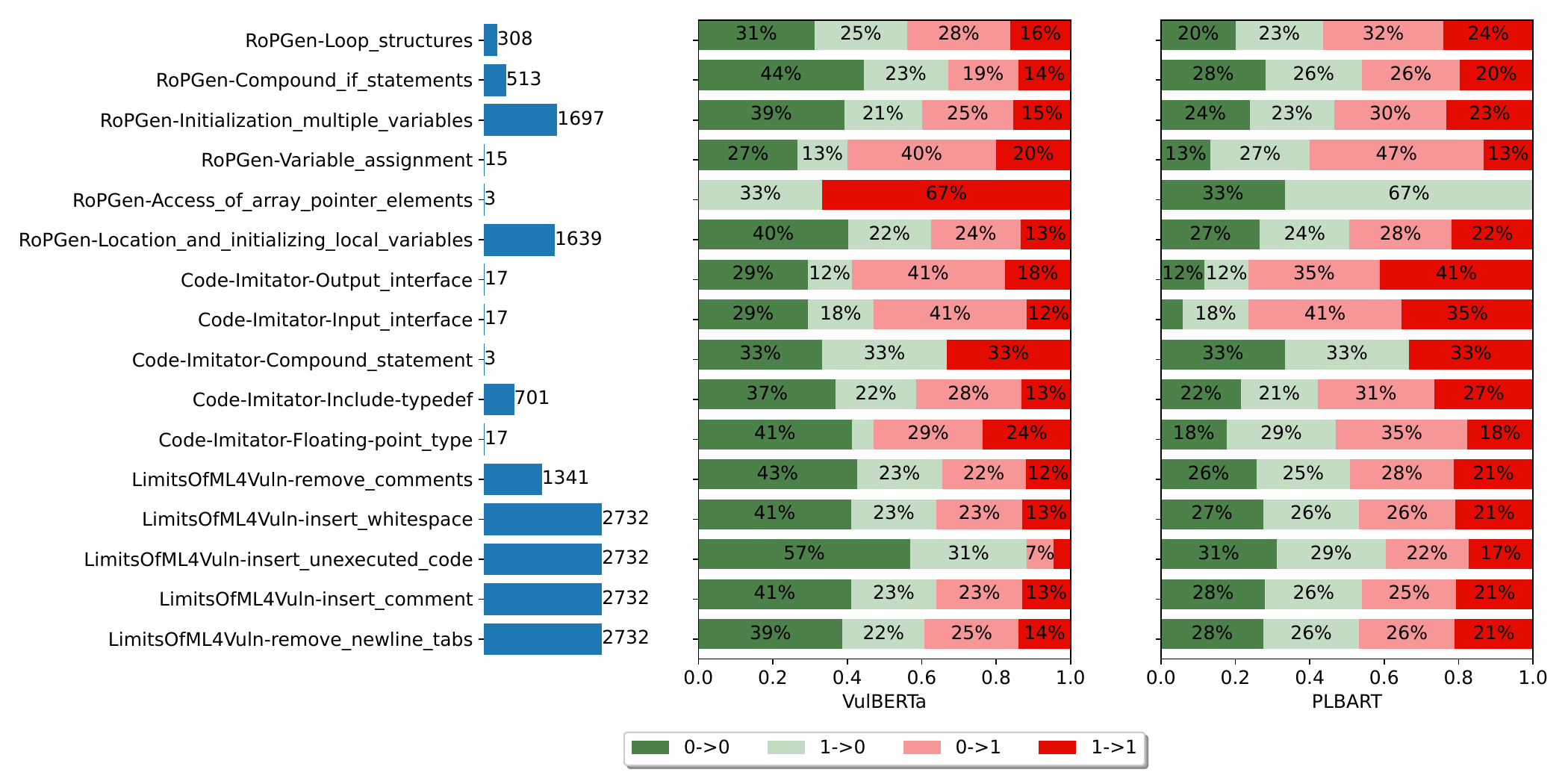}
\caption[]{RQ1: Illustration of the frequency each transformation can be applied (left) and resulting impact on the predicted labels for VulBERTa and PLBART (center and right). 
For example, ``0-\textgreater1'' means that prediction for the original function was 0 and is 1 for the transformation. Here, $0$ means the function is non-vulnerable and $1$ means vulnerable.}
\label{fig:compare}
\end{figure*}

In this research question, we address how the \appliedTransforms semantic preserving transformations impact the predictions made by two LLMs (VulBERTa, PLBART). 
For this purpose, we apply each transformation to every function in the Devign test set, if they are applicable.

First, we consider how frequently each transformation can be applied.
Figure~\ref{fig:histogram} illustrates the number of transformations applied to each of the 2732 functions in the Devign test set. 
We can observe that, each function can be transformed 6.3 times on average, with values ranging from 4 to 12.
Only three functions received 12 transformations, \changed{which means that we obtain 12 transformed variants on which a single transformation has been applied to.}

Figure~\ref{fig:compare} (leftmost chart) shows how often each of the transformation can be applied. 
We were able to apply four out \appliedTransforms transformations to each of the functions in the Devign test set. 
These modify whitespaces, comments and dead code.
Transformation to the initialization of variables, as provided by RopGen, can be applied to 60\% of the functions.
The remaining transformations can only be applied rarely, with some being applied only three times.
Moreover, Figure~\ref{fig:compare} presents the predictions made on the transformed functions and whether they change the prediction of the original function.
\changed{In particular, we illustrate whether a predicted label remains at 1 (``1-\textgreater1''), changes from a 1 to a 0 (``1-\textgreater0''), changes from a 0 to a 1 (``0-\textgreater1'') or remains at 0 (``0-\textgreater0'').
$1$ indicates that a function was labeled as vulnerable, while a label of $0$ indicates a non-vulnerable function.}
Here, the ratios of the four groups varies between the respective transformations with 51\% of labels \changed{not changing after applying transformations} (either ``0-\textgreater0'' or ``1-\textgreater1'').
The behavior differs between VulBERTa and PLBART, as can be seen in the $insert\_unexecuted\_code$ (LimitsOfML4Vuln).
The majority of functions labeled as non-vulnerable by VulBERTa remain unchanged, while PLBART is more likely to predict them as vulnerable.
We continue to investigate the ability of transformations to correct defect detection tools in RQ2.

\begin{mdframed}[style=mystyle]
\noindent
\textbf{Answer to RQ1:}
Semantic-preserving transformation change predictions made by the two LLMs in 49\% of the cases. These can either change vulnerable to non-vulnerable predictions (25\%) or vice-versa (24\%).
\end{mdframed}

\subsection{RQ2: Defect Detection Ensembles}
In the second research question, we investigate the use of semantic-preserving transformations to enhance the testing of defect detection tools.
For this purpose, we employ three different ensemble strategies (Section~\ref{section:ensemble}) based on combinations of transformations and LLMs.
Table~\ref{table:rq2} compares the accuracy of the original defect detection tools (VulBERTa, PLBART) with the accuracy of the ensemble approaches.
Unfortunately, we observe that our ensemble strategies have not been able to increase the accuracy over the original VulBERTa model while performing better than PLBART.

Among the two strategies for ties, breaking ties at $0$ always performs better than breaking ties at 1, meaning that the a predicted label of 0 (non-vulnerable) is preferred. 
This can be explained by the fact that the majority class is $0$, and thereby more functions should be labeled as non-vulnerable to achieve a high accuracy.
Weighted approaches seem beneficial when dealing with a higher number inputs (i.e., the 16 transformations). %
However, the highest accuracy of ensemble approaches is achieved by averaging the two LLM predictions, without considering transformations.
A reason for this could be that results are difficult to generalize. %

Generally speaking, averaging prediction probabilities performs better than simply relying on labels for voting.
The advantage of averaging is that the certainty of predictions is taken into account.

\begin{mdframed}[style=mystyle]
\noindent
\textbf{Answer to RQ2:}
Our investigated ensemble strategies were not able to increase the accuracy of defect detection tools. 
\end{mdframed}

\begin{table}
\centering
\caption{RQ2: Performance of ensemble strategies.}
\label{table:rq2}
\begin{adjustbox}{max width=.92\columnwidth}
\rowcolors{1}{}{Gray2}
\begin{tabular}{lllr}
\toprule
 &  & VulBERTa & PLBART \\
 \midrule
 & Original & 64.71 & 61.79 \\
 & CodeXGLUE & 64.75  & 63.18 \\ \midrule
Data ensemble & Majority - Ties 0 & 51.43 & 48.13  \\
 & Majority - Ties 1 & 51.39 & 47.91 \\
 & Average & 52.49 &  50.84\\
 & Weighted - Labels & 52.34  &  54.28\\
 & Weighted - Probability & 52.12 & 59.19 \\ \midrule
Model ensemble & Majority - Ties 0 & \multicolumn{2}{c}{58.86} \\
 & Majority - Ties 1 & \multicolumn{2}{c}{54.54} \\
 & Average & \multicolumn{2}{c}{62.52} \\
 & Weighted - Labels & \multicolumn{2}{c}{61.79} \\
 & Weighted - Probability & \multicolumn{2}{c}{61.75} \\\midrule
Data and model & Majority - Ties 0 & \multicolumn{2}{c}{52.12} \\
 & Majority - Ties 1 & \multicolumn{2}{c}{51.79} \\
 & Average & \multicolumn{2}{c}{52.12} \\
 & Weighted - Labels & \multicolumn{2}{c}{60.03} \\ 
 & Weighted - Probability & \multicolumn{2}{c}{60.61} \\ \bottomrule
\end{tabular}
\end{adjustbox}
\end{table}

\subsection{Replication Difficulty}
\label{section:replication}

We have found \numTransforms transformations from \pubSemantic publications (Section~\ref{sec:transformations}).
Among these, 9 publications did not share their code, which complicates a replication.
The remaining \pubRelevant studies consider different programming languages, which limits their applicability.
In the end, we considered four publications in detail, as they are applicable for C/C++ code and implemented several transformations. 
39 of the transformations provided by the four publications were applicable to the Devign dataset (e.g., changes to functions).
We were able to re-implement and confirm the semantic-preserving behavior for \appliedTransforms of these (i.e., by manually checking up to 20 results for each transformation), none of which were part of NatGen~\cite{chakraborty2022:natgen}.

We found that transformations can be task-dependent and therefore not always easy to transfer.
For example, Quiring et al.~\cite{quiring2019:misleading} applied transformations for authorship disguise. 
At times, this requires the presence of a source file (file to be transformed) and target file (style information). 
Given that we only have a single function at hand, we were not able to use all available transformations.

We have also observed some edge cases, which do not seem to have been considered in the original publications.
This can be explained by the different datasets being used.
For instance, functions in the Devign dataset can contain assembly code which the variable renaming function of RoPGen did not consider.
Another example can be seen in the following code snippet which outlines an erroneous transformation of moving variable definitions inside control statements:
\begin{lstlisting}[language=Python,xleftmargin=0.5cm]
// Before transformation
unsigned i;
for (i = 0; i < 10; i++)
    foo();
for (i = 0; i < 10; i++) 
    bar();
    
// After transformation
for (unsigned i = 0; i < 10; i++) 
    foo();
for (i = 0; i < 10; i++) 
    bar();
\end{lstlisting}
As can be seen from this example, the second for loop, after transformation, is referring to the variable $i$, however it is only defined in the loop and therefore causes errors due to accessing an undefined variable.

Lastly, the scope of transformations can be different. For example, some transformations are applied on a file-level, while we are focused on functions.
Transformations that require files, such as the addition of imports, the use of main functions, or the presence of multiple functions in general, could not be implemented.
This shows that one needs to be careful when transferring semantic-preserving transformations from one application to another. 

\begin{mdframed}[style=mystyle]
\noindent
\textbf{Replicability:} Existing semantic-preserving transformations are difficult to replicate due to missing implementations, different programming languages and application scenarios as well as potential incorrect behavior when dealing with edge cases.
\end{mdframed}

\section{Conclusions}
\label{section:conclusion}
\noindent
We investigated the use of semantic-preserving transformations for mutating source code to enhance the testing of defect detection tools. %
We first searched and found \pubSemantic publications which implemented such transformations and picked four to implement for our experiments.
Overall, it appears challenging to reuse existing methods, due to differences in application scenarios, programming languages or errors due to edge cases.
In the end, we were able to successfully use \appliedTransforms out of 39 available transformations and used them in ensemble approaches for defect detection.
While we provide insights on the usefulness of transformations and replication difficulty, we were not able to improve the accuracy of the considered defection detection tools with mutations and ensembles.

Future work efforts can strive to extend the empirical evaluation of transformations for defect detection. 
This includes additional LLMs, datasets and transformations (e.g., error-inducing transformations).
Moreover, one could apply more than a single transformation for each of the functions~\cite{pour2021:searchbased,quiring2019:misleading}.

The \numTransforms transformations we found stem from \pubSemantic publications, some of which do not provide implementations. 
Therefore, we believe a consolidated framework of semantic-preserving transformation, for different modalities (function or file-level) and programming languages, could benefit the community.
One important consideration of such a framework is the provision of proofs to guarantee the correctness of transformations.

\section{Data Availability}
We share our results and all transformations applied to the Devign dataset to enable a replication and verification of our results: \url{https://figshare.com/s/f9b93d9d549f316b1e5d}.

\section*{Acknowledgment}
\noindent
This work is supported by the Research Council of Norway through the secureIT project (IKTPLUSS \#288787), and by the European Union through the Horizon Europe Marie Sk\l{}odowska-Curie Actions (\#101151798).

\printbibliography

\clearpage{}%

\end{document}